
\input harvmac
\noblackbox

\def \bb  {{\bar \b }}

\def \ra {\rightarrow}

 \def \k1 {{1\over
k}}  \def \ov { \over }

\def \a {\alpha}
\def \b {\beta}

\def \ln {{\rm \ ln \  }}

\def \ch {{\rm cosh \ }}
\def \th {{\rm tanh  }}
\def \l {\lambda}
\def \1p {{1\over  \pi }}
\def \2p {{{1\over  2\pi }}}
\def \4p {{ {1\over 4 \pi }}}
\def \8p {{{1\over 8 \pi }}}
\def \P^* { P^{\dag } }
\def \p {\phi}

\def \m {\mu }
\def \n {\nu}

\def\g {\gamma}
\def \r {\rho}
\def \k {\kappa }

\def \s {\sigma}
\def \t {\theta}

\def \Gmn {G_{\m\n}}
\def \Bmn {B_{\m\n}}
\def \fourth {{{1\over 4}}}

\def \e#1 {{{\rm e}^{#1}}}

\def \eq#1 {\eqno {(#1)}}

\def \ov {\over }


\def \p {\phi}

\def \s {\sigma}

\def \r {\rho}

\def \l {\lambda}
\def \m {\mu}
\def \g {\gamma}
\def \n {\nu}

\def \e#1 {{{\rm e}^{#1}}}

\def \vp {\varphi}

\def \P {\Phi}
\def\np {  Nucl. Phys. }
\def \pl { Phys. Lett. }
\def \mpl { Mod. Phys. Lett. }
\def \prl { Phys. Rev. Lett. }
\def \pr  { Phys. Rev. }

\def \cmp { Commun. Math. Phys. }
\def \ijmp { Int. J. Mod. Phys. }
\baselineskip8pt
\Title{\vbox
{\baselineskip8pt{\hbox{CERN-TH.6970/93}}{\hbox{hep-th/9308042}}{\hbox{(revised)}} }}
{\vbox{\centerline { On field redefinitions and exact solutions}\vskip2pt
 \centerline{in string theory }
}}
\vskip  -1 true pt
\centerline{A.A. Tseytlin\footnote{$^{*}$}{\baselineskip8pt
On leave  from Lebedev Physics
Institute, Moscow, Russia.
e-mail: tseytlin@surya3.cern.ch and tseytlin@ic.ac.uk} }
\smallskip
\centerline {\it Theory Division, CERN}
\centerline {\it
CH-1211 Geneva 23, Switzerland}
\centerline  {and}
\centerline {\it Theoretical Physics Group, Blackett Laboratory}
\centerline {\it Imperial College}
\centerline {\it London SW7 2BZ, U.K.}
\vskip 20 pt
\centerline {\bf Abstract}
\medskip
\baselineskip10pt
\noindent
String backgrounds associated with  gauged $G/H$ WZNW models
in general depend non-trivially on  $\alpha'$. We note,
however, that there exists  a  local covariant $\a'$-dependent
field redefinition  that relates the exact metric-dilaton
background corresponding to the $SL(2,R)/U(1)$ model to
its  leading-order form ($D=2$ black hole).  As a
consequence, there exists a  `scheme' in which the string
effective equations have the latter as an exact solution.
However, the corresponding   equation for the tachyon
(which, like other Weyl anomaly equations, has scheme-dependent
form) still contains corrections of all orders in $\alpha'$. As
a result, the  string `probes'   still   feel the
$\alpha'$-corrected  background.
The field redefinitions we  discuss  contain
the dilaton terms in the metric transformation law.
We also comment on  exact forms of the duality transformation
in different `schemes'.

\vskip 80 pt
\noindent
{CERN-TH.6970/93}
\smallskip
\noindent
 {August 1993}
\Date { }
\noblackbox
\baselineskip 20pt plus 2pt minus 2pt


\lref \nem {D. Nemeschansky and S. Yankielowicz, Phys.Rev.Lett. 54(1985)620;
54(1985)1736(E).}
\lref \bpz {A.A. Belavin, A.M. Polyakov and A.B. Zamolodchikov, \np
B241(1984)333. }
\lref \efr {S. Elitzur, A. Forge and E. Rabinovici, \np B359 (1991)
581;
 G. Mandal, A. Sengupta and S. Wadia, Mod. Phys. Lett. A6(1991)1685. }

\lref \bcr {K. Bardakci, M. Crescimanno and E. Rabinovici, \np
B344(1990)344. }
\lref \Jack {I. Jack, D.R.T.  Jones and J. Panvel,  \np B393(1993)95. }
\lref \tse { A.A. Tseytlin, \pl B264(1991)311. }

\lref \frts {E.S. Fradkin and A.A. Tseytlin, Phys.Lett. B158(1985)316;
Nucl.Phys. B261(1985)1. }
\lref \mplt {A.A. Tseytlin, \mpl A6(1991)1721. }
\lref\bn {I. Bars and D. Nemeschansky, \np B348(1991)89.}
\lref \shif { M.A. Shifman, \np B352(1991)87.}
\lref \dvv  { R. Dijkgraaf, H. Verlinde and E. Verlinde, \np B371(1992)269. }
\lref \kniz {  V. Knizhnik and A. Zamolodchikov, \np B247(1984)83. }

\lref \witt { E. Witten, \cmp 92(1984)455.}
\lref \wit { E. Witten, \pr D44(1991)314.}
\lref \anton { I. Antoniadis, C. Bachas, J. Ellis and D.V. Nanopoulos, \pl B211
(1988)393.}
\lref \ts  {A.A. Tseytlin, \pl B268(1991)175. }
\lref \kir {{E. Kiritsis, \mpl A6(1991)2871.} }

\lref \shts {A.S. Schwarz and A.A. Tseytlin,  preprint Imperial/TP/92-93/01
(1992). }
\lref \bush { T.H. Buscher, \pl B201(1988)466.  }

\lref \plwave { D. Amati and C. Klim\v cik, \pl B219(1989)443; G. Horowitz and
A. Steif, \prl 64(1990)260.}

\lref\bsft { I. Bars, preprint USC-91-HEP-B3. }

\lref\bs { I. Bars and K. Sfetsos,  \pr D48(1993)844. }
\lref \tsw  { A.A. Tseytlin,\np B399(1993)601.}
\lref \call {C.G. Callan, D. Friedan, E. Martinec and M.J. Perry,
Nucl. Phys.B262(1985)593.}
\lref \met {R.R. Metsaev and A.A. Tseytlin, Phys.Lett. B191(1987)354;
Nucl.Phys. B293(1987)385.}
\lref \aps { S. De Alwis, J. Polchinski and R. Schimmrigk, \pl B218(1989)449. }
\lref \gny  { M.D. McGuigan, C.R. Nappi and S.A. Yost, \np B375(1992)421. }
\lref \py  {  M.J.  Perry and E. Teo, preprint DAMTP R93/1 (1993); P. Yi,
preprint CALT-68-1852
(1993). }
\lref \witten { E. Witten,  Commun. Math. Phys. 144(1992)189. }
\lref \sfts {K. Sfetsos and A.A. Tseytlin, preprint CERN-TH.69.. ,  to appear.}
\lref \gin  { P. Ginsparg and F. Quevedo, \np B385(1992)527.}
\lref \sft    {K. Sfetsos, \np B389(1993)424.}
\lref \napwi  { C. Nappi and E. Witten, \pl }
\lref \st {K. Sfetsos and A.A. Tseytlin, preprint CERN-TH.6969/93.}
\lref \curci { G. Curci and G. Paffuti, \np B286(1987)399.   }
\lref \ajj {R.W. Allen, I. Jack and D.R.T. Jones, Z. Phys. C41(1988)323. }
\lref \ttt {A.A. Tseytlin. \pl B178(1986)349.}
\lref \hul {C. Hull and P.K. Townsend, \pl B191(1987)115;
D. Zanon, \pl B191(1987)363; D.R.T. Jones, \pl B191(1987)363;
S. Ketov, \np B294(1987)813.}
\lref \hult {C.M. Hull and P.K. Townsend, \np B301(1988)197.}
\lref \jaj {I. Jack and D.R.T. Jones,  \pl B200(1988)453; \np B303(1988)260.}
\lref \tsey {A.A. Tseytlin, \pl B176(1986)92; \np B276(1986)391.}
 \lref \grwi {  D. Gross and E. Witten, \np B277(1986)1. }
\lref \myers { R. Myers, Phys.Lett. B199(1987)371;
I. Antoniadis, C. Bachas, J. Ellis and D. Nanopoulos,
\pl B211(1988)393. }
\lref \parall { T.L. Curtright  and C.K. Zachos, \prl 53(1984)1799;
S. Mukhi, \pl B162(1985)345; S. de Alwis, \pl B164(1985)67. }
\lref \tsmpl {A.A. Tseytlin, \mpl A6(1991)1721.}
\lref \scsc  {J. Scherk and J. Schwarz, \np
 B81(1974)118.}

\lref \calg { C.G. Callan and Z. Gan, \np B272(1986)647. }
\lref \tss {A.A. Tseytlin, \pl B178(1986)34. }
\lref \fri  {D.H. Friedan, \prl 45(1980)1057; Ann. Phys. (NY) 163(1985)318. }
\lref \tsss { A.A. Tseytlin, \ijmp A4(1989)1257. }
\lref \osb { H. Osborn, Ann. Phys. (NY) 200(1990)1.}
\lref \jackk {I. Jack, D.R.T. Jones and D.A. Ross, \np B307(1988)130.}
\lref \jackkk {I. Jack, D.R.T. Jones and N. Mohammedi, \np B332(1990)333.}
\lref \shts  {A.S. Schwarz and A.A. Tseytlin, \np B399(1993)691.}

\newsec { General remarks on  field redefinition ambiguity }

String  theory is effectively a non-local theory  when
represented in terms of the standard local fields.
  If ones uses the couplings $\vp^i = (\Gmn, \Bmn,\p,...)$ of a $2d$
$\s$-model  to
parametrise string backgrounds then the  (string tree level)  effective  action
$S(\vp^i)$ \scsc\
or the corresponding equations  ($\s$-model conformal invariance conditions
$\bb^i=0$)   contain terms
of
 all orders in derivatives of the fields multiplied by powers of  $\a'$,
$$ \bar \b^i = \k^{ij} {\del S \ov \del \vp^j} =  {\bar\b}^i_1 (\vp)  + \a'
{\bar \b}^i_2(\vp)  + ...
=0  \ .  \eq{1} $$
Since ${\bar\b}^i_1$  are  of second order in derivatives,
solving (1) perturbatively in $\a'$  we find  $$
\vp^i_* = \vp^i_1 + \a' \vp^i_2 + ... \ \   , \ \ \ \ \
 \vp^i_2 = -\big[\big({\del{\bar \b}^i_1 \ov \del \vp^j}\big)^{-1} {\bar
\b}^j_2\big] ({\vp_1} )
\ \  , \
... \ \   ,  \eq{2} $$ where higher-order corrections $\vp^i_2, \vp^i_3, ..., $
are   in general
{\it  non-local}, being expressed in terms of the leading-order solution
$\vp_1^i$.

 If the only information one uses  is  an  on-shell string $S$-matrix, then
the effective action  is defined  modulo local field redefinitions (since they
do not change the
$S$-matrix) \tsey\grwi\
$$ \vp^{i\prime}= \vp^i +  \a' T^i_1 (\vp) + ... \ . \eq{3} $$
Here $T^i_n$  are local combinations of the fields  containing  $2n$
derivatives.
A similar ambiguity  (or `scheme dependence')
is present   in the $\bb^i$-functions.\foot {This is not exactly  the same as
the scheme ambiguity
present in the standard $\b$-functions; the redefinitions  we consider are  of
more   general type,
 see below. } The redefinitions (3) should preserve the
structure of $S$ and $\bb^i$, i.e. should respect their  symmetries  (like
general covariance    and
$\p \ra \ \p +$ const).
 The transformation  (3) changes the structure of higher-order terms in $S$ and
in $\bb^i$, e.g.
$$ \bb^{i\prime}_2= {\bar \b}^i_2 + {\del{\bar \b}^i_1 \ov \del \vp^j}  T^j_1 \
,\   \ ... \  \  ,
\eq{4}  $$ and, as a result, of the solution (2). The solution of the
transformed equations will, in
general, remain a non-local functional of $\vp_1^i$.

It may happen that there  exists  such a special  representative  (or a  `
special scheme')
 in the class of equivalent (redefinition-related) actions $S$  or conformal
anomaly coefficients $\bb^i$  for which
all higher-order corrections in (1) vanish on a particular leading-order
solution $\vp^i_1$ (so that
this leading-order solution is actually an exact one).   Then the solution  for
a generic choice of
$S$ (or in a generic `scheme')  will be  a function of $\a'$ given by a {\it
local} redefinition (3)
of the special scheme solution $\vp^i_1$. A  well-known example is provided by
the
parallelizable spaces \parall\ (in particular, by group spaces corresponding to
WZNW theories \witt) :
there exists  a special scheme  \hul\met\
    in which  each  term  $\bb^i_n$ in (1)  vanishes,  being evaluated on
$\vp^i_*=\vp^i_1$.
The solution in a general scheme will be a  `deformation'  of $G_{*\m\n} $ and
$ B_{*\m\n} $
by local $\a'$-dependent terms (i.e. by $a_1\a' R_{*\m\n } + ... $ and $a_2 \a'
D^\l_*
H_{*\l\m\n}+...$).\foot {For group spaces this amounts  to a rescaling of
$\Gmn$ and $\Bmn$ by
functions of $\a'R $ $ (= \ha Dc_G /k)$.  This changes the   distribution of
contributions to the
central charge \nem\ $\ C= Dk/(k+ \ha c_G)$ coming from different loop orders
in $\bb^\p$. In
particular, one can find a scheme in which  $\Gmn$ is effectively rescaled by
$k/(k+\ha c_G)$ (which
amounts to changing $\a'$ from $1/k$ to $1/(k+\ha c_G)$)  so that the whole
central charge  comes
only from  the `one-loop' $O(\a')$ correction to $\bb^\p$. }

In the previous example the dilaton is constant in the special scheme and
remains constant in all
other schemes as well.  The situation becomes more subtle once the dilaton is
non-constant for a
leading-order solution.
 This is the case that we will be interested in  below.  We shall
ignore all other fields in the theory,   except the metric and the dilaton.
There exists a scheme
\call\met\ in which there  is only one   ($R^2_{\l\m\n\s}$)  $\a'^2$-term  in
the effective  action
$$  S =    \int d^{D}x \sqrt {\mathstrut G} \ {\rm e}^{- 2
\phi} \lbrace {1\ov 6}  (D-C)  - {1\ov 4}  \alpha^{\prime} [R    +  4
(\partial_{\mu} \phi)^{2}]
  - {1 \over 16} \a'^2   R_{\m\n\l\k}^2 +  O(\a'^3)   \rbrace \ , \eq{5} $$
where $C$ is a total central charge.
The corresponding  equations  are linear combinations of the  $\bb^i$-functions
$$
{\bar \b}^G_{\m\n} = R_{\m\n}  + 2D_\m D_\n \p  + \ha \a' R_{\m\l\r\s}R_\n^{\
\l\r\s} + O(\a'^2)
= 0 \ , \eq{6} $$
  $$ {\bar \b}^\p =   {1\ov 6} (D-C)   + \a' [ - \ha  D^2 \p + (\del \p)^2]
+ { 1\ov 16}\a'^2 R^2_{\l\m\n\r} + O(\a'^3)  =0 \ .   \eq{7} $$
 In a general scheme
the $\a'$-terms in (6) and $\a'^2$-terms in (5),(7) will depend on a number of
free parameters,
which enter  the most general local field redefinition \tsey\met\foot{One can
ignore the term
$D_\m D_\n \p$ in (8) since it can be eliminated by   a coordinate
transformation along  $D^\m \p$.}
 $$ G'_{\m\n} = G_{\m\n} +  \a'[ d_1 R_{\m\n}  + d_2 \del_\m \p \del_\n \p  +
 G_{\m\n} ( d_3 R  + d_4 D^2\p + d_5 (\del \p)^2)]   +
O(\a'^2) \ , \ \ \ \eq{8} $$
$$\p'=\p + \a'[ d_{6} R  + d_{7} D^2\p + d_{8} (\del \p)^2] + O(\a'^2)  \ .
\eq{9} $$
Since $\Gmn$ and $\p$ appear  on an equal footing in the string effective
action,   it  seems  natural
to consider the most general local covariant  redefinitions that mix them.
Note that  the transformations of $\Gmn$ that involve the dilaton
do not have an interpretation  as corresponding to a change of a
renormalisation scheme in the
standard $\s$-model $\b$-functions. However,  since the dilaton is a coupling
constant of  a
$\s$-model defined on a curved $2d$ background \frts,  such transformations
should  have
a  meaning   of  changes of a  `scheme'  corresponding to  the  conformal
anomaly coefficients
$\bb^i$.

The transformations (8),(9) are to be understood within the $\a'$ perturbation
theory.
Consider, for example,  the simplest case
of the linear dilaton background \myers
 $$ \Gmn = \eta_{\m\n} \ , \ \ \ \  \p = \p_0 + n_\m x^\m \ , \ \ \ \  {1\ov 6}
(D-C)   + \a'n^2
=0 \ . \eq{10} $$
 Making the redefinition (8), i.e.
 $$ G'_{\m\n} = \eta_{\m\n}    +  \a'  \del_\m \p \del_\n \p
F_1(\a'\del\p\del\p)  + \eta_{\m\n} F_2(\a'\del\p\del\p)    = c_1 \eta_{\m\n} +
c_2 \a'n_\m n_\n \
\ ,   $$   one can significantly alter the structure of the
metric (for example,  change  its  signature or even  make it degenerate)
unless $ |\a' n^2|\ll 1$.
Unless $C=D$ (i.e.  unless  $n^2=
\eta_{\m\n}n^\m n^\n=0$)   one cannot,  strictly speaking,
use the $\a'$-perturbation theory  since  here the derivative of the dilaton
(or dilaton `momentum')
is not small compared with  $\a'^{-1/2}$. One can, however, consider  the
linear dilaton background
as a part of  more general solution  where $\a'n^2$ is a free parameter.

With similar  clarifications,  we would like  to suggest that
 the transformations  (8),(9)  are
 perfectly admissible. The effective action $S$ and the $\bb^i$-functions
 should have  a background-independent meaning:  the use
of the flat space string  $S$-matrix  (or perturbation theory near any other
particular background)
is a technical tool for
establishing  the structure of $S$. Having determined it,   one may relax  all
 assumptions (such $C=D$) about a background used   and consider all  solutions
of the same
background-independent equations on an equal footing.   Moreover,  one is free
to choose {\it any}
representative in a class of equivalent actions (related by field redefinitions
of the general type
(8),(9))  and use it as a  starting  point.  Different choices of the
actions correspond to different off-shell extensions of string theory. Unless
one  has an extra
principle for fixing a particular one,  one is   free  to  make arbitrary
redefinitions  like
(8),(9).

\newsec{Exact `$SL(2,R)/U(1)$' solution   and its relation to the
leading-order   $D=2$
black-hole background}
Let us now apply the above  general remarks  to  the  case of a  particular
$D=2$ solution of  the
bosonic  string theory. Equations (6),(7) have the following leading-order
(Euclidean `black-hole')
solution \efr\wit  $$ ds^2 =  dx^2 + a^2 (x)  \ d{\theta}^2  \   , \ \ \  \ \
a^2 (x) = {\th}^2 nx
+ O(\a') \ , \ \ \eq{11}  $$ $$  \p  =  \p_0  -  {\ln \ch}  nx
 + O(\a') \ ,  \ \ \  \   {1\ov 6} (D - C)   + \a'n^2 =0 \ . \eq{12} $$
For this background
$$ R = {4n^2\ov {\rm cosh}^2nx}\ , \   \    (\del \p)^2 = n^2 {\th}^2 nx
 \  , \ \
  - \ha D^2 \p + (\del \p)^2 =   \fourth R +  (\del \p)^2 = 4n^2   \ . \eq{13}
$$
Since $\bb^\p$ in (7) is constant on a solution of $\bb^G=0$ (6)  \curci,  one
may  keep $n$ as a
free parameter  by not imposing the  central-charge  constraint (7).\foot{As
was mentioned above, to
avoid the  question  about the  validity of perturbation theory in $\a'$ once
(7) is imposed, we may
assume that  this background is a part of a more general solution so that the
total $C$ and $D$ are
such that $\a' n^2$ is effectively small. To make $\a'n^2$  a continuous
parameter
 we may introduce, for example,   a linear dilaton background  ($ - n_0 t$) in
the
direction  of an additional time-like coordinate so that   $n^2 -n^2_0=0$. }
Solving the two-loop metric equation in the  standard scheme, where it has the
form (6),
one finds the following corrections \ts\foot{The two-loop contributions  to the
central-charge
equation cancel out.
 In the standard scheme used in (6),(7)
, the  total  contribution to $\bb^\p$ comes from the asymptotic
large-distance value of the  `one-loop' dilaton  term $(\del \p)^2$.}
$$ a^2(x) = {\th}^2 nx + 2\a'n^2 {\th}^4 nx + O(\a'^2)\ ,
\eq{14} $$
$$\p =  \p_0  -  {\ln \ch}  nx
 + \ha \a'n^2   {\th}^2 nx  + O(\a'^2) \ . \eq{15} $$
 It was found
   that in the three-  \ts\ and four-  \Jack\ loop  approximation there exists
such a scheme that the  resulting solution reproduces  the $\a'$-expansion of
the
following  generalisation of (11),(12) \dvv:
$$a^2(x) = { {\th}^2 nx \ov 1 \ - {\ p \ } {\th}^2 nx }=  {(1+ 2\a'n^2) {\rm
tanh}^2 nx
 \ov 1 + 2 \a'n^2 ({\rm
\cosh}\  nx)^{-2}}
\  , \ \ \  p\equiv {2\a'n^2\ov 1+2\a'n^2} \ ,  \eq{16} $$ $$ \p = \p_0  -
{\ln \ch }  nx   -
\fourth  \ln [ 1 \ - \ p \ {\th}^2 nx ]$$ $$ = \p'_0  -  {\ln \ch }  nx   -
\fourth  \ln [ 1 +   2
\a'n^2 ({\rm \cosh}\  nx)^{-2}]
  \ . \eq{17} $$
This suggests that  there exists a `standard' scheme  in which (16),(17) is an
exact solution,
in agreement with   the  derivation of this background from the
$SL(2,R)/U(1)$ coset conformal field theory  \dvv\   and  the corresponding
gauged WZNW model
\tsw\bs.\foot{If one imposes the condition that the total central charge $C=2 +
6\a'n^2$ in (7),(12)
is equal to that of the  $SL(2,R)/U(1)$  coset model, $C= 3k/(k-2) -1$, then
the
parameter $\a'n^2$  becomes  related to $k$:  $\a'n^2= 1/(k-2)$,      $\
p=2/k$. }

The schemes for the $\b$-functions considered in \ts\Jack\ were  related  to
the  `dimensional
regularisation plus minimal subtraction' one
 by  local  redefinitions  that  {\it did not} include the dilaton-dependent
terms.
The result of \ts\  was that it  is in  {\it such}  schemes  that  the
leading-order  solution
(11),(12) is necessarily modified by the $\a'$ corrections at higher loop
orders  (with the two-loop
terms   having the  unambiguous form (14),(15)).

The main observation we would like to make  here  is that the exact background
(16),(17)
can be represented as a local {\it dilaton-dependent}  field redefinition of
the leading-order
one   (11),(12).  Given that  (16),(17) is  the exact solution in the
`standard' scheme, this implies
that  there exists a  `non-standard' scheme in which the
leading-order  solution (11),(12)  is, in fact,  an {\it exact solution}  to
all orders in $\a'$.
In  such  a scheme each of the  higher-loop contributions to   $\bb^G$ vanishes
 separately on the
background (11),(12).

Consider the following field redefinition in $D=2$ (cf. (8),(9))
$$    G'_{\m\n} = G_{\m\n} +  \a' \del_\m \p \del_\n \p  F_1(\a'R, \a'\del
\p\del \p )
             +  G_{\m\n} F_2(\a'R,\a'\del \p\del \p )   \ , \eq{18}   $$
$$ \p' = \p  +  F_3(\a'R,\a'\del \p\del \p ) \ ,  \eq{19}  $$
where the functions $F_s$ are  given  by  power series  of their arguments.
Taking
$$ F_1= -2 + O(\a')\ , \ \ \ F_2 = 2\a' (\del \p)^2 + O(\a'^2) \ , \ \ \
F_3= - {1\ov 8}  \a' R + O(\a'^2) \ , \eq{20} $$
one finds   that (18),(19)  computed for (11),(12) reproduce the
$\a'$-corrections  (14),(15) to the
leading-order solution. This means that making the redefinition (18),(19),(20)
in
 (6),(7) one will find the two-loop equations  corresponding to a `scheme' in
which the
leading-order  solution (11),(12) remains also the solution in the $\a'$-
approximation.

The exact form of  the redefinition  that transforms (11),(12) into the
$\a'$-dependent
 background  (16),(17)
is given by (18),(19),    with\foot{It is interesting to note that the leading
order
correction term in (18),(21)  $-2\a'\del_\m \p \del_\n \p $ is precisely the
same that appears
(along with the dilaton term) in the determinant produced by integration over
the gauge field
 in the corresponding gauged WZNW action. This determinant was computed in
\shts.
The term $\sim \int \del_\m \p \del_\n \p \del_a x^\m \del^a x^\n$
in the resulting $2d$ effective action can be ignored in the  leading one-loop
approximation but
contributes at the two-loop level (there will be also other two-loop terms; the
form of all such
terms  will depend on a  computational scheme used). }   $$  F_1 =    { -2 \ov
1  + \ha \a' R  }
       \ ,  \ \ \
\  F_2 =     { 2\a'(\del \p)^2 \ov 1 + \ha \a' R   }  \ , \ \ \ \  F_3=  -
\fourth  \ln ( 1  +
{\textstyle {1\ov 2}} \a' R  )        \ .   \eq{21}      $$
This is readily checked using (13).
As a consequence, there exists a scheme (found by applying (18),(19),(21) to
the `standard' scheme)
in which (11),(12) is an exact solution.
The main point is that the transformation between (11),(12) and (16),(17) can
be represented
in a local  {\it background-independent} form.
Note that the transformation for the inverse $D=2$ metric
$$    G^{\m\n \prime } = G^{\m\n} f_1 +  \a'D^\m \p  D^\n \p  f_2
            \  \ ,  $$  $$  f_1 = {  1 + \ha \a' R \ov
  1 +  \ha \a' ( R + 4 \del_\m \p \del^\m \p) } \ , \ \ \
  f_2 = {  2 \ov
  1 +  \ha \a' ( R + 4 \del_\m \p \del^\m \p) } \ ,  \eq{22} $$
takes a simple form on the leading order solution  (11),(12) (denominators in
(22) become constant,
cf. (13))  but still contains   all  higher order terms
 when represented in the background-independent form (22).

Since  the transformation (18),(19),(21) is invariant
 under   constant shifts of $x$ (e.g.
$x \rightarrow x +\ha  i \pi /n$),  it also transforms  the dual
to the leading-order solution  (11),(12)
 (with cosh replaced by sinh)  into the  exact background `dual' to (16),(17).
As a consequence,  in the  `non-standard' scheme in which (11),(12) is an exact
solution,
its dual in the sense of the usual leading-order form of duality
$$\tilde G_{\t\t} = G_{\t\t}^{-1}, \  \ \ \
 \tilde \p = \p - \ha \ln G \ , \eq{23}
$$
is also an exact solution.
Therefore, the duality transformation  also has its leading-order form in such
a scheme.
To find  the  exact ($\a'$-corrected \tsmpl) form that the  duality
transformation  has in the
`standard' scheme,   one is to invert the transformation  (18),(19),(21),  then
apply the
leading-order  duality   (23),
 and finally transform the result  again  with the help of (18),(19),(21).
 The  product of the three
transformations will  be given by a power series in $\a'$ and will relate, in
particular,
(16),(17) to  its exact dual (with $x \rightarrow x +\ha  i \pi /n$).

\newsec {Tachyon equation in a background}

One is  left with the question about  the physical implications of the
existence
of a scheme in which  the leading-order background (11),(12) is an exact
solution.
While all higher string modes have zero condensates  for this solution   one
should look at
linearised equations for their perturbations in order to understand how the
string
 feels the background (i.e. one should perturb the conformal theory by marginal
operators as
`probes').  In general, these  equations (or corresponding Weyl anomaly  $\bar
\beta$-functions)
are linear differential equations with background field dependent coefficients
 which are also
{\it scheme dependent}.   It may happen  that in the scheme where the
background fields  have
semiclassical form the equations for the `probes' still contain
$\a'$-corrections of all orders.
This, in fact, what happens in the  $D=2$  case considered above:
while the equation for the tachyon has a  simple form in the  scheme where
  (16),(17) is the solution, it becomes complicated in the new scheme.
Being evaluated on the  corresponding metric-dilaton solution, the tachyon
equation takes  of course  the same $\a'$-dependent form in both schemes
(coinciding with the $(L_0 +
{\bar L_0} ) T =2T$  equation of  conformal theory  approach \dvv ).

The  tachyon  ${\bar\b}$-function has the following  general
structure\foot {  Here we  consider the case of arbitrary dimension $D$ and
include also the
dependence on the field strength $H$  of a possible antisymmetric tensor
background.} \fri\calg\tss\
(see also \tsss\osb)  $$ {\bar\b}^T = - \gamma T  + W^\m \del_\m T - 2T \ , \
\eq{24} $$
where $ \g$ is the scalar anomalous dimension operator
and  $W^\m$ is the `diffeomorphism vector' \tss
$$ \gamma =  \Omega_2^{\m\n} D_{(\m} D_{\n)}
 + \sum^\infty_{n=3} \Omega_n^{\m_1 ... \m_n}  D_{(\m_1} ... D_{\m_n)}  \ ,
\eq{25} $$
$$ \Omega_2^{\m\n}  = \ha \a' G^{\m\n} + p_1 \a'^2 R^{\m\n}  + p_2 \a'^2
H^{\m\a\b}H^\n_{\ \a \b}   +
p_3 \a'^2 R^\m_{\a\b\g} R^{\n\a\b\g}  + ... \ , \ \  \eq{26} $$ $$
\Omega_3^{\m\n\r}  = \a'^4 q_1 D_\a R_{\b \ \ \g}^{ \ \m\n} R^{\a\b\g\r}  + ...
\ , \ \
\Omega_4^{\m\n\r\l }= \a'^4 s_1 R^{\m\a\b\n} R^{\r\ \ \ \l}_{\ \a\b} + ... \ ,
\eq{27} $$
$$ W_\m = \a'\del_\m \p + M_\m (G,H)\ ,  $$ $$  M_\m = t_1\a'^2 \del_\m
(H^2_{\a\b\g} )
 + t_2 \a'^2 D^\n (H_{\m\a\b}H_\n^{\ \a \b} )  + t_3 \a'^3 \del_\m(
R^2_{\a\b\g\r})  + ... \ . \eq{28}
$$ The coefficients $p_1,p_2, ...$ are obviously scheme-dependent.
In the  dimensional regularisation / minimal subtraction
scheme $p_1=0$ \calg\tss,  the coefficients $p_2,t_1,t_2$ can be inferred  from
\met\osb, $p_3 = {3\ov 16}, \ t_3= {1\ov 32} $ \jackk\ and the four-loop
coefficients
in $\Omega_2$ and $q_1,s_1, ...$ were found in \jackkk.
 Thus if $H=0$ the leading order (one-loop) form of the tachyon equation
$$  {\bar\b}^T = - \ha \a' D^2  T  +  \a' D^\m \p  \del_\m T - 2T =0 \ , \
\eq{29} $$
is not modified by the two-loop corrections in this scheme.

 Let us now recall  that  the exact $D=2$ background  (16),(17) was found in
\dvv\ by
identifying the $(L_0 + {\bar L_0} ) T =2T$  equation of  $SL(2,R)/U(1)$ coset
conformal
theory with  the {\it leading order} form of the  tachyon equation (29).
The  consistency of such identification then demands  that the scheme in which
(16),(17)
is the exact solution of the metric-dilaton  $\bar\b^i =0$  equations  (which
exists
according to \ts\Jack\  in the four-loop approximation)
must be the one in which the tachyon $\bar\b$-function  maintains its
leading-order form (29).
This, in fact, is  easy to check in the  three loop approximation: the
$R^2$-term  appearing  in
(26)  in the minimal subtraction scheme \jackk\ is redefined away by the  same
transformation
(from the minimal subtraction scheme to the  `standard' one  corresponding  to
(16),(17))  that was
found   in \ts.\foot { Let us note that the  expressions for the coefficients
$a_i$  in
terms of $c_i$ in eq.(23) of ref. \ts\   had  erroneously extra factors of 2. }
We expect that similar  statement is  true   in  four (and higher) loop
approximation (to get
rid of higher derivative terms in (25) one will need also to redefine the
tachyon field).
As a result,   one can define the scheme in which (16),(17) is an exact
solution as the one in which
${\bar\b}^T$   takes  the simple form (29).

In conclusion, let us
 emphasize that the  existence of   a     local, covariant  and
background-independent  transformation,  which,  like  (18),(19),(21),
relates the  leading-order
solution to the exact one,  is quite  non-trivial (as we have noted above, a
generic   exact solution
is non-local,  being expressed (2) in terms of the leading-order one).
Equivalently, the existence of
a `scheme' in which  a leading-order solution is  exact to all orders   should
be a property of  only
a very  special class of solutions.
Given that the group spaces   and now the simplest $SL(2,R)/U(1)$ coset model
background
belong to such a class,  one may  conjecture that this is actually true for all
solutions  that correspond to  $G/H$ coset models, i.e. for all  $\s$-model
backgrounds  that
originate from gauged WZNW theories.

 However, this fact may have rather limited importance  since the quantum
string modes
propagating in these  backgrounds will  always  `feel'  the $\a'$-corrected
solutions:
 the corresponding equations will be  equivalent to the $(L_0 + {\bar L}_0)T =
n T$ - equations
of the coset $G/H$ conformal field theories which depend non-trivially on the
level $k$
in all (bosonic) cases with  the subgroup $H\not=1$. This  suggests that  the
`standard'
  scheme    in which  the
tachyon equation has  the simple form (29)  and the background  fields contain
$1/k$  corrections of
all orders  is   a `preferrred' one being directly related to the conformal
field theory
 interpretation.

\bigskip
\bigskip
{\bf Acknowledgements}

\noindent
I am grateful to K. Sfetsos for a collaboration on related issues  \st\
 and to Al.B. Zamolodchikov  for  a   stimulating discussion.
I would like to thank H. Osborn for raising the  important question  about the
form of the tachyon
$\beta$-function and   I. Jack  and J. Russo   for  comments.
I   acknowledge  also  a   support of SERC.

\vfill\eject
\listrefs

\end